\documentclass[amsmath,amssymb,aps,prb,twocolumn,superscriptaddress]{revtex4-2}
\usepackage{graphicx}
\usepackage{bm}

\begin{document}
\title{Topological edge states and amplitude-dependent delocalization in quasiperiodic elliptically geared lattices}
\affiliation{Department of Physics, University of Michigan, Ann Arbor, MI 48109, USA}
\affiliation{Key Lab of Advanced Optoelectronic Quantum Architecture and Measurement (MOE), School of Physics, Beijing Institute of Technology, Beijing 100081, China}
\affiliation{Department of Mathematics and Statistics, University of Massachusetts, Amherst, MA 01003, USA}
\affiliation{Department of Mechanical Engineering, Seoul National University, Seoul 08826, Republic of Korea}

\author{Shuaifeng Li}
\affiliation{Department of Physics, University of Michigan, Ann Arbor, MI 48109, USA}
\author{Di Zhou}
\affiliation{Key Lab of Advanced Optoelectronic Quantum Architecture and Measurement (MOE), School of Physics, Beijing Institute of Technology, Beijing 100081, China}
\author{Feng Li}
\affiliation{Key Lab of Advanced Optoelectronic Quantum Architecture and Measurement (MOE), School of Physics, Beijing Institute of Technology, Beijing 100081, China}
\author{Panayotis G. Kevrekidis}
\affiliation{Department of Mathematics and Statistics, University of Massachusetts, Amherst, MA 01003, USA}
\author{Jinkyu Yang}
\affiliation{Department of Mechanical Engineering, Seoul National University, Seoul 08826, Republic of Korea}

\begin{abstract}
We present a class of mechanical lattices based on elliptical gears with quasiperiodic modulation and geometric nonlinearity, capable of exhibiting topologically protected modes and amplitude-driven transitions. Starting from a one-dimensional chain of modulated elliptical gears, we demonstrate the emergence of localized edge states arising from quasiperiodic variation in the gears' moments of inertia, analogous to the topological edge modes of the Aubry-Andr\'e-Harper model. Under increasing excitation amplitude, the system undergoes a nonlinear transition, where edge localization breaks down and energy delocalizes into the bulk. By coupling multiple such chains with varying modulation phase, we construct a two-dimensional lattice in which the phase acts as a synthetic dimension. This structure supports topological wave propagation along the synthetic dimension. Nonlinearity again induces a breakdown of topological states, leading to complex, amplitude-dependent wave propagation. We further propose a numerical continuation approach to analyzing the periodic orbits and their linear stability, effectively discovering the boundary of the basin of bounded motion and detecting the occurrence of delocalization under certain excitation amplitudes. Our results reveal that elliptical geared systems offer a passive, amplitude-dependent platform for exploring topological phenomena and synthetic dimensionality in mechanical metamaterials.
\end{abstract}

\maketitle

\section{Introduction}\label{sec1}
The field of mechanical metamaterials has rapidly advanced toward systems where geometry and topology interplay to yield unconventional, disorder-immune mechanical response~\cite{ni2023topological,xin2020topological}. Central to this progress is topological band theory, first established in electronic~\cite{gilbert2021topological} and photonic systems~\cite{ozawa}, which now enables robust defect-immune edge states across diverse mechanical platforms including gyroscope lattices~\cite{nash2015topological,wang2015topological}, soft matter~\cite{li2018observation,zhou2018PRL,zhou2019PRX,li2019valley,li2021topological}, origami structures~\cite{miyazawa2022topological,li2023geometry,li2024emergence,li2024topological,li2025demand}, and isostatic networks~\cite{rocklin2017transformable,mao2018maxwell,tang2024fully}. Recent breakthroughs extend these concepts beyond physical spatial dimensions through synthetic dimensions, such as parametric degrees of freedom~(e.g., temporal modulation or spatially varying phases), which emulate higher-dimensional topology in lower-dimensional mechanical systems~\cite{rosa2019edge,xia2020topological,wang2023smart,riva2020adiabatic,riva2020edge}. This approach introduces phenomena such as topological pumping and adiabatic state transfer, promising for wave control beyond physical spatial constraints.

Among these mechanical platforms, geared metamaterials offer a distinctive paradigm where rotational coupling and gear geometry encode topological physics. By tailoring parameters such as ellipticity and moments of inertia, these systems achieve programmable stiffness landscapes and unconventional mechanical behavior. For example, gear clusters with stiffness gradients achieve continuously tunable elastic properties and shape morphing~\cite{fang2022programmable,mo2024continuously}, while coupled translational-rotational designs realize topological floppy modes and mechanical analogues of Weyl semimetals~\cite{meeussen2016geared}. Particularly relevant to our work, elliptical gear chain exhibits topological zero-frequency modes, with recent studies revealing their potential for nonlinear topological transitions~\cite{ma2023nonlinear}. Despite these advances, critical gaps persist in the previous works. First, synthetic dimensions remain unexplored in geared systems, limiting access to higher-order topological phenomena in lower-dimensional systems. Second, nonlinear dynamics, especially amplitude-dependent state transitions, are poorly characterized despite geometric nonlinearity being intrinsic to elliptical gear kinematics. Third, while amplitude-driven transitions occur in systems such as nonlinear spring-mass systems and origami~\cite{li2024topological,rosa2023amplitude,zhou2022NC}, they remain unexplored in systems with topological zero modes where unique edge-dependent dynamics may emerge.

Here, we address these challenges by introducing quasiperiodic elliptically geared lattices. By spatially modulating the moment of inertia of elliptical gears in a 1D chain, we impose controlled quasiperiodicity mimicking the Aubry-Andr\'e-Harper~(AAH) model~\cite{harper1955single,harper1955general}. We demonstrate that this geared chain supports robust topological edge modes with localized energy distribution. Building upon this 1D foundation, we construct a 2D synthetic lattice by coupling parallel 1D chains with a phase modulation, enabling propagation of edge states along a single boundary. Furthermore, we demonstrate how geometric nonlinearity intrinsic to elliptical gear interactions triggers amplitude-dependent topological edge states. While small excitations preserve edge localization, increasing the driving amplitude triggers a transition to delocalized bulk states. Through proposed numerical continuation method and Floquet stability analysis, we map the boundaries of basin of bounded motion of periodic orbits, revealing how initial conditions determine localized edge states or delocalized bulk states.

This transition highlights the fundamentally dynamic nature of topological states in geared systems, whose response can be affected not only by geometry and boundary conditions but also by excitation amplitude. Our findings not only reveal the dynamical properties of a system with zero-frequency modes, but also demonstrate elliptical geared metamaterials as a versatile platform to study topological mechanics, synthetic dimensions, and nonlinear dynamics.

\section{1D quasiperiodic elliptically geared lattices}\label{sec2}
We begin our exploration with a one-dimensional mechanical chain composed of paired elliptical gears. Unlike conventional circular gears with constant radii, elliptical gears introduce the geometric nonlinearity, enabling unprecedented dynamical properties. The elliptical gears in the chain are connected via meshing teeth and rotated around pivots on their focal points. Here we focus on the rotation around the right focal point~[$F_{2}$ in Fig.~\ref{fig:fig1}(a)]. In order to conform to the AAH model and to maintain gear engagement during rotation, we spatially modulated the moment of inertia of individual gears according to  $I_{n}=I_{0}+\mu I_{0}\sin(2\pi n\xi+\eta)$, resulting in an effective modulation of the on-site potential. Here, $\mu$, $\xi$ and $\eta$ represent the modulation amplitude, modulation frequency and modulation phase~(also the synthetic dimension of the system in the following section). This modulation introduces spatially varying effective stiffness and effectively maps the system onto a mechanical analog of the AAH model. In our work, we select $N=28$ elliptical gears, where each gear is labeled by an index $n$~($\{n\in\mathbb{Z}|0\leq n\leq 27\}$), each of which has one rotational degree of freedom $\theta_{n}$. A rational $\xi$ results in a periodic pattern of the chain while an irrational $\xi$ leads to a quasiperiodic pattern of the chain.

To model the dynamics of this system, the Lagrangian of the chain can be written as:
\begin{equation}
    \label{equ:equ1}
    L=T-U=\frac{1}{2}\sum_{n}(I_{n}\dot{\theta}_{n}^2-k_{n}l_{n}^2),
\end{equation}
where $k_{n}$ is the elastic constant~(reflecting gear stiffness) and $l_{n}$ is the sliding distance~(reflecting teeth deformation) between gear $n$ and gear $n-1$. The sliding distance is given by $l_{n}=s_{e}(\theta_{n})-s_{-e}(\theta_{n+1})$, where $s_{e}(\theta)$ is the arc length of the contacting point traveling along the ellipse with eccentricity $e$ and rotation angle $\theta$. Specifically, in the small-eccentricity limit with $e\ll1$, the arc length is approximated as $s_{e}(\theta)=a(\theta-e\sin{\theta})$~\cite{ma2023nonlinear}.

By leveraging the Euler-Lagrange prescription, we obtain the equation of motion for the $n$-th gear in the form:
\begin{equation}
    \label{equ:equ2}
    \frac{d}{dt}\left(\frac{\partial{L}}{\partial{\dot{\theta}_{n}}}\right)-\frac{\partial{L}}{\partial{\theta_{n}}}=0,
\end{equation}
\begin{equation}
    \label{equ:equ3}
    I_{n}\ddot{\theta}_{n}+\tau_{n}-\tau_{n-1}=0,
\end{equation}
where subscripts $n$ and $n-1$ represent the torque functions from the adjacent gears. The detailed derivation of Eq.~\eqref{equ:equ3} can be found in the Appendix~\ref{appendixA}. 

\subsection{Linear spectrum}\label{sec2_1}
For small angular displacements, the system behaves linearly and we linearize Eq.~\eqref{equ:equ3} as follows:
\begin{equation}
    \label{equ:equ4}
    I_{n}\ddot{\theta}_{n}+\alpha\theta_{n}+\beta\theta_{n-1}+\gamma\theta_{n+1}=0.
\end{equation}
The spectrum of vibrational modes can be obtained by diagonalizing the corresponding effective dynamical matrix, as we now explore.

\begin{figure}[h]
    \centering
    \includegraphics[width=0.5\textwidth]{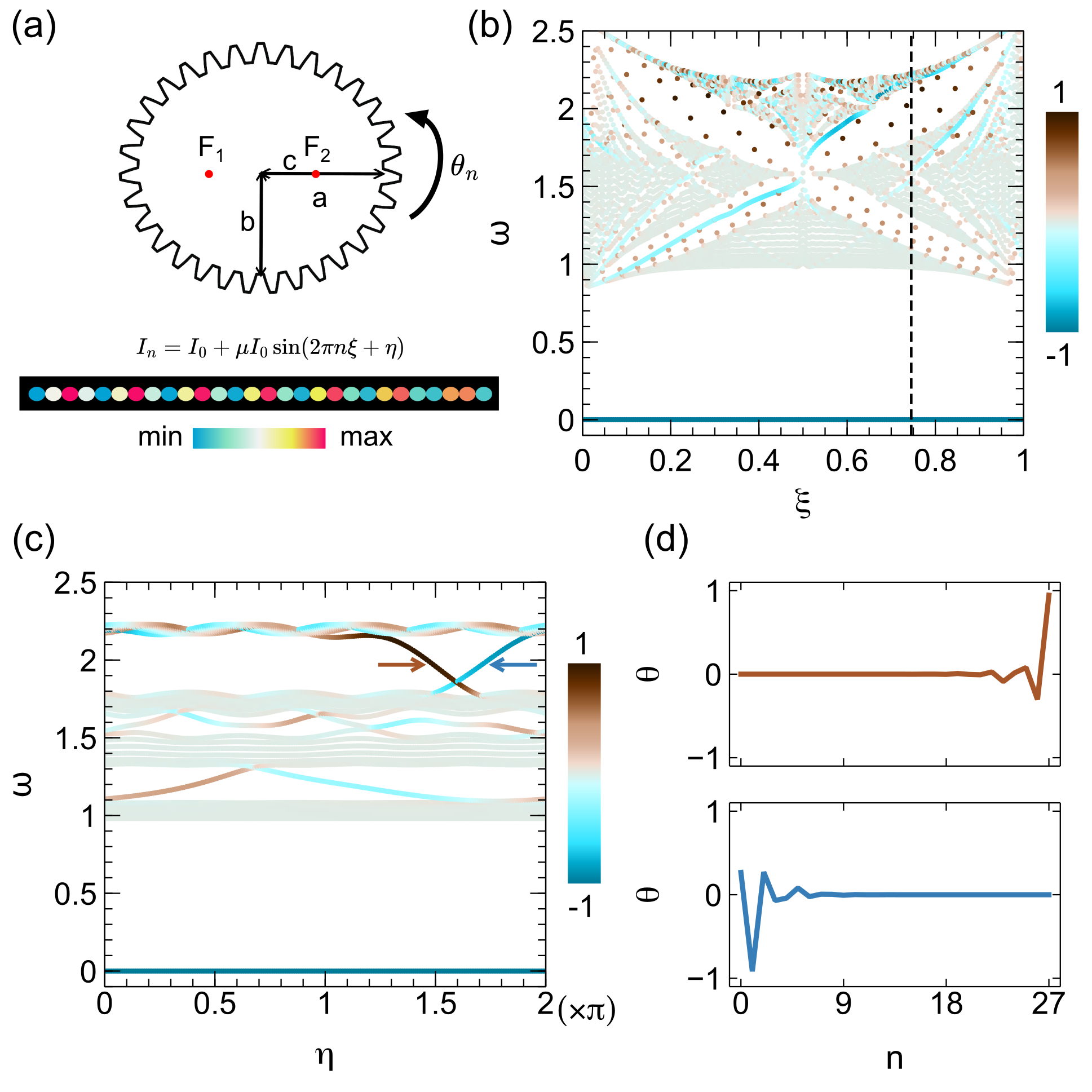}
    \caption{(a) The schematic of an elliptical gear. $a$, $b$ and $c$ represent lengths of major axis, minor axis and center to focus. $F_{1}$ and $F_{2}$ represent two foci. The counterclockwise rotation is defined as the positive direction. An elliptically geared chain with spatial modulation is shown below. The moment of inertia of each gear follows $I_{n}=I_{0}+\mu I_{0}\sin{(2\pi n\xi+\eta)}$, which is encoded by color.
    (b) The calculated eigenspectrum of a finite chain with $28$ elliptical gears with free boundary conditions as a function of $\xi$. The black dashed line represents $\xi=\frac{\sqrt{5}}{3}$.
    (c) The calculated eigenspectrum of a finite chain with $28$ elliptical gears with free boundary conditions as a function of $\eta$. The color in (b) and (c) represents the localization index, indicating the degree of the energy localization. Two arrows indicate two points where $\omega=2$.
    (d) Two eigenmodes~($\theta$) of the elliptical gear chain, corresponding to two points in (c).
    \label{fig:fig1}
    }
\end{figure}

In our work, we use the dimensionless geometric parameters: major axis $a=1$, eccentricity $e=0.5$, initial moment of inertia $I_{0}=1$ and modulation amplitude $\mu=0.4$, and the dimensionless mechanical parameter: $k_{n}=1$ for the elastic constants. In Fig.~\ref{fig:fig1}(a), we visualize the geometry of a gear and a gear chain with modulated moment of inertia represented by color.

To study the topological properties of the elliptical geared chain, we show the calculated spectra. In Fig.~\ref{fig:fig1}(b), the eigenspectrum is plotted as a function of the modulation frequency $\xi$ with the free boundary condition on both sides of the geared chain, when the modulation phase $\eta=0$. We observe a structure of dispersion relation known as Hofstadter butterfly~\cite{hofstadter1976energy}, indicating both the spectral fractality and nontrivial band gaps with nonzero gap Chern numbers~\cite{prodan2019k}. The color represents the localization index (LI), which is the product of the inverse participation ratio (IPR) and the center of mode (CoM), characterizing the localization of energy~\cite{li2024topological}. It is defined as:
\begin{equation}
    \label{equ:equ5}
    LI=IPR\times CoM
\end{equation}
with
\begin{equation}
    \label{equ:equ6}
    IPR=|(\bm{\Theta}^{\circ 2})^{\circ 2}|
\end{equation}
\begin{equation}
    \label{equ:equ7}
    CoM=\bm{w}^{T}\bm{\Theta}^{\circ 2}
\end{equation}
Here, $\bm{\Theta}$ is the eigenvector containing the rotational degree of freedom. The weighting vector $\bm{w}$ is given by $\bm{w}=[-N/2, \cdots, -1, 1, \cdots, N/2]^{T}$. $\circ$ denotes the Hadamard product. The introduction of LI indicates that if the eigenmode is skewed toward the left~(right) boundary, LI is negative~(positive). The resulting spectrum reveals a set of discrete bulk bands separated by gaps that host localized modes at the edges of the chain. Besides, the zero-frequency modes can also be observed, which support the localization of energy at the left edge~(soft edge).

According to the color-encoded spectrum in Fig.~\ref{fig:fig1}(b), topological edge states with the energy localized on either boundary can be found in the band gap when $\xi$ is in a certain range. We focus on the quasiperiodicity introduced by an irrational number $\xi=\frac{\sqrt{5}}{3}$ in our work to further explore the topological states when $\eta$ varies from $0$ to $2\pi$, as shown in Fig.~\ref{fig:fig1}(c). Notably, from $1.5\pi$ to $1.7\pi$, there are topological states in the band gap with gap Chern number $+1$ around $\omega=2$. We examine the eigenmodes when $\omega=2$ indicated by arrows in Fig.~\ref{fig:fig1}(c). The eigenmodes displayed in Fig.~\ref{fig:fig1}(d) clearly exhibit the topological edge states with localized energy on either the soft edge or rigid edge.

\subsection{Nonlinear response}\label{sec2_2}
While the linear response of the elliptically geared chain reveals the presence of topologically protected edge states, the richness of this gear system emerges when nonlinear effects are taken into account. The nonlinear response originates from the elliptical geometry of the gears, particularly under large angular displacements. In this regime, coupling terms can no longer be treated as constants, and the system undergoes a qualitative change in its dynamical behavior.
\begin{figure}[h!]
    \centering
    \includegraphics[width=0.5\textwidth]{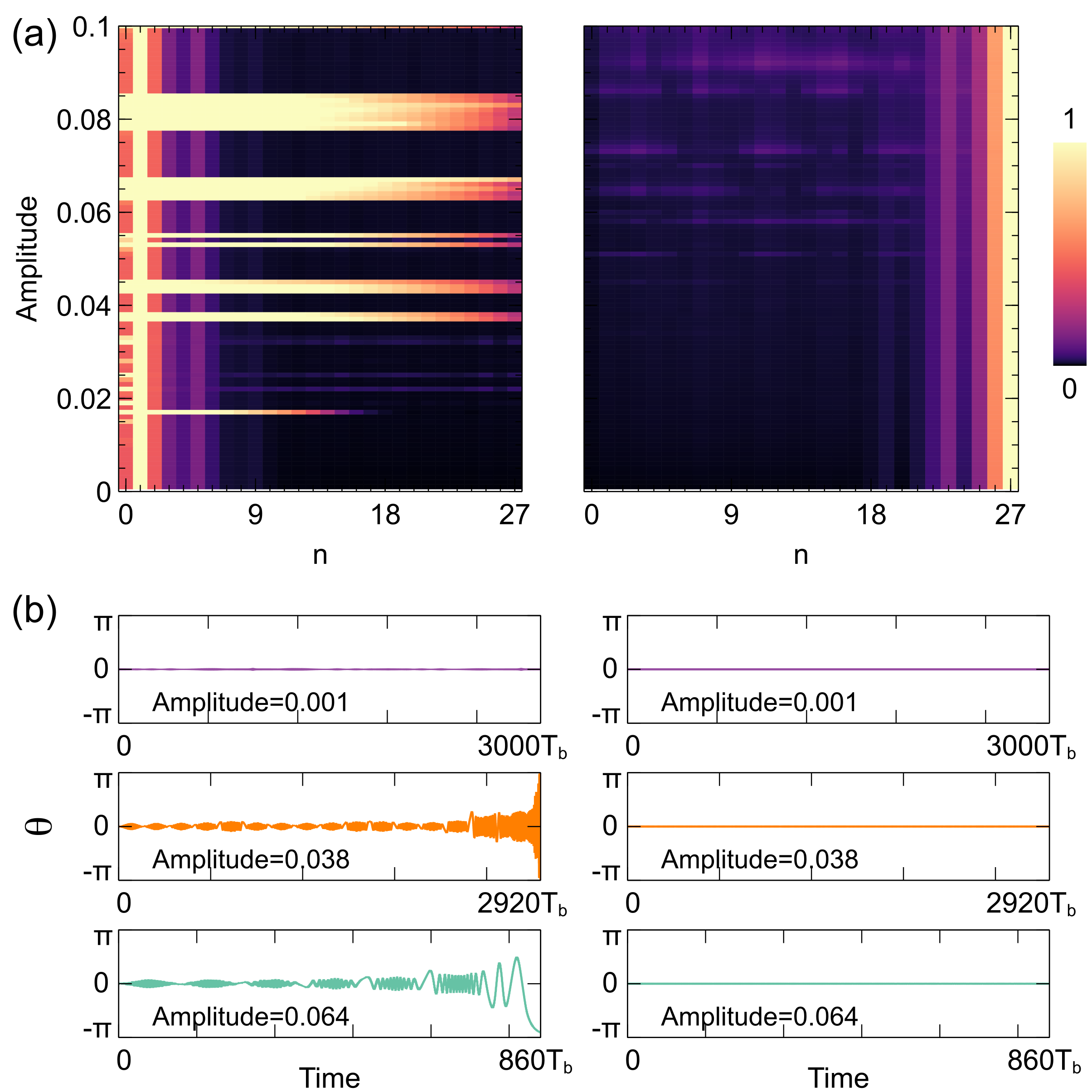}
    \caption{(a) The left panel shows the RMS of $\theta$ as a function of excitation amplitude from the soft side. The right panel shows the RMS of $\theta$ as a function of excitation amplitude from the rigid side.
    (b) The left three graphs show the evolution of $\theta$ of the gear on the soft side over time under different excitation amplitudes. The right three graphs show the evolution of $\theta$ of the gear on the rigid side over time under different excitation amplitudes.
    \label{fig:fig2}
    }
\end{figure}

To investigate this change, we excite the gear at the edges~(either the soft edge or the rigid edge) using a harmonic excitation~($\omega=2$) with different amplitudes over a certain period of time $T=3000T_{b}$ ($T_{b}=\frac{2\pi}{\omega}$) and examine the resulting motion throughout the chain. Fig.~\ref{fig:fig2}(a) illustrates this transition. The first panel shows the normalized root mean square~(RMS) of the field distribution $\theta$ as a function of excitation amplitude, when the chain is excited from the soft edge. In the low-amplitude regime~(near the linear regime), energy remains localized near the soft edge, consistent with the presence of a topological edge state. However, as the excitation amplitude increases, the response delocalizes into the bulk, indicating a breakdown of the topological protection and a transition to a more extended mode. Importantly, the nonlinear delocalization is neither gradual nor a threshold-dependent behavior. There are exceptions where even large-amplitude excitation cannot trigger the nonlinear delocalization. Instead, the edge modes still remain. The transition process is further visualized by checking the rotation angle $\theta$ of the boundary gear at the soft edge in Fig.~\ref{fig:fig2}(b). Under different excitation amplitudes $0.001$, $0.038$ and $0.064$, we observe a transition time point under large excitation amplitudes when the loss of edge localization happens, which exhibits time-dependence with its inverse correlation to excitation amplitude. At an amplitude of $0.064$, the transition occurs at $860T_{b}$ earlier than at $0.038$ ($2920T_{b}$). In Supplementary Video 1, we show the time-domain simulation where the rotation of gears is clearly visualized.

However, when the excitation is applied at the rigid edge, as shown in the second panel of Fig.~\ref{fig:fig2}(a), energy remains localized near this rigid edge regardless of the excitation amplitude. The right panel of Fig.~\ref{fig:fig2}(b) illustrates the time-dependent rotation of the rigid-edge gear under varying excitation amplitudes. In the Supplementary Video 2, we show the time-domain simulation under different excitation amplitudes, where transition waves are not excited. This stark contrast between edge behaviors stems from the instability of the soft mode at the soft edge. Qualitatively, for small displacements, the system’s linear response dominates, and energy remains confined to the excited boundary~(either soft edge or rigid edge). However, as the amplitude increases at the soft edge, the soft mode destabilizes under certain amplitudes, causing the gear at the soft edge to rotate by $\pm\pi$. This triggers a transition wave that propagates toward the rigid edge, effectively converting the rigid edge into a soft edge. The process does not stabilize. Instead, the transition wave cycles repeatedly. This transition can be clearly observed in Supplementary Video 1. Nevertheless, when excitation is at the rigid edge, this instability and the consequent transition wave fail to arise, preserving the localized energy distribution (Supplementary Video 2). 

\section{Delocalization analysis under nonlinear excitations}\label{sec3}
The preceding section on 1D quasiperiodic elliptically geared lattices has demonstrated amplitude-dependent transitions where topologically protected edge states break down and delocalize into the bulk. Moreover, these transitions exhibit edge dependence. The results suggest the potential delocalized dynamics under specific excitations, requiring a systematic analysis in order to corroborate the relevant dynamical observations.

To investigate this, we develop a method for finding a branch of periodic orbits $X_{j}$~($j$ is the index of excitation amplitudes) across excitation amplitudes in our lattices, leveraging the topological states in the linear regime and numerical continuation. Inspired by the work of~\cite{marin1996breathers} and its application to numerous fields~\cite{FLACH20081}, including on breathers in mechanical models for DNA~\cite{slade2010stability}, we consider the topological states starting from the linear regime, rather than a decoupled nonlinear setting, as is the case in the widely used anti-continuous limit~\cite{RSMacKay_1994}. A standard Newton method then traces a branch of precise (to a prescribed accuracy) periodic orbit solutions across excitation amplitudes. Specifically, we begin in the linear regime by applying a small excitation at the frequency of the topological states~($\omega=2$), and use the resulting solution at a given time as the initial seed. Once the periodic orbit solution for this seed is found, numerical continuation incrementally increases the amplitude parameter from the linear regime towards the nonlinear regime of larger values. This iterative Newton-based process tracks periodic orbits into the desired excitation amplitude regime. The details can be found in Appendix~\ref{appendixB}.
\begin{figure}[h!]
    \centering
    \includegraphics[width=0.5\textwidth]{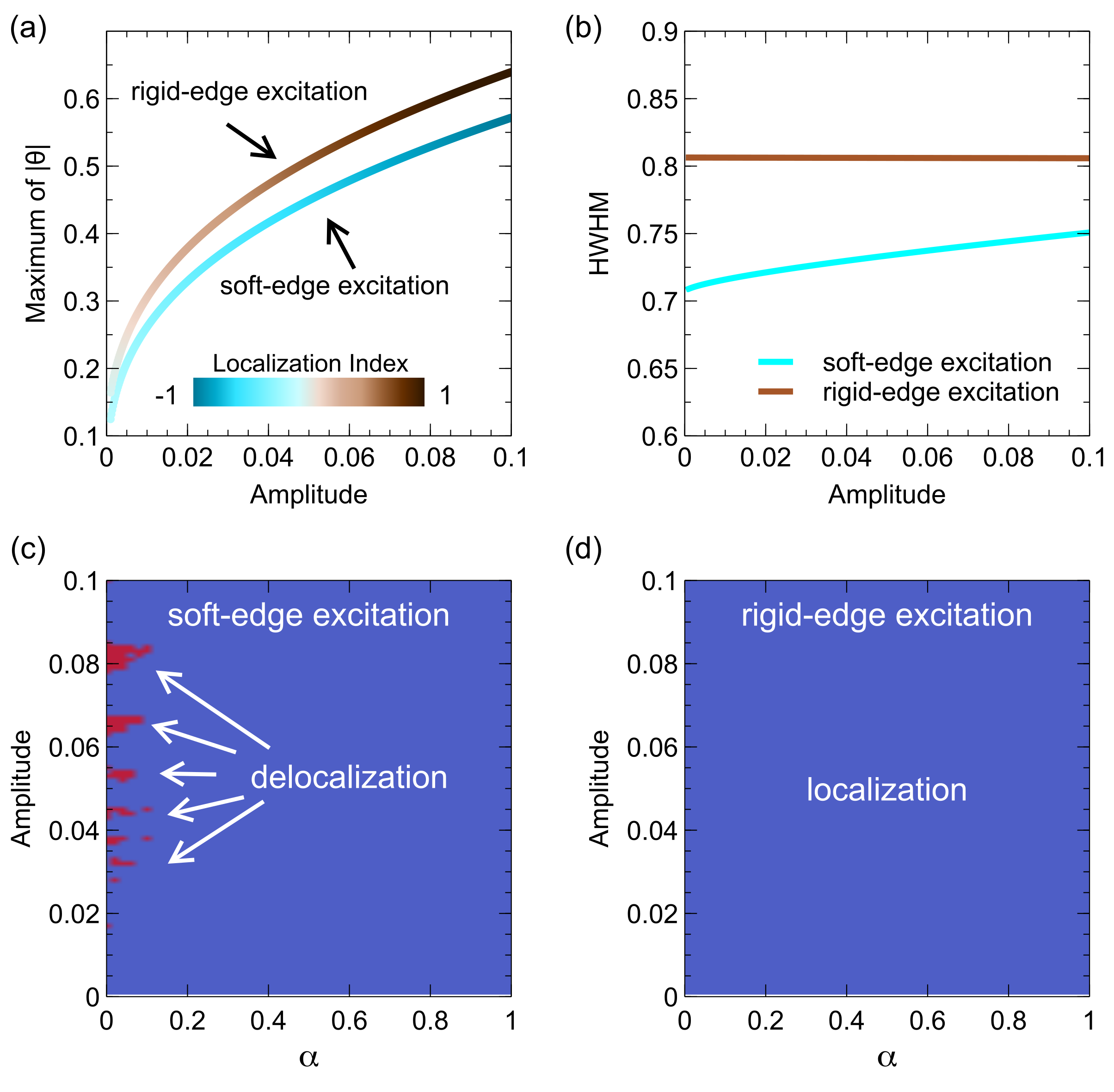}
    \caption{(a) The maximum of $|\theta|$ of the periodic orbits as a function of the excitation amplitudes. The color represents the localization index.
    (b) The HWHM of the periodic orbits as a function of the excitation amplitudes.
    (c) (d) When the initial condition varies linearly from the periodic orbit solution to zeros, the localized dynamics is indicated by blue and the delocalized dynamics is indicated by red. The case for soft-edge excitation is shown in (c) and that for rigid-edge excitation is shown in (d).
    \label{fig:fig3}
    }
\end{figure}

Once a branch of periodic orbits for our system is established via the numerical continuation method, we assess the linear stability of each solution, by examining the eigenvalues of the associated monodromy matrix~\cite{FLACH20081}. First, we check the localization. Fig.~\ref{fig:fig3}(a) shows the maximum of $|\theta|$ of periodic orbits versus excitation amplitude, along with LI. For both soft-edge and rigid-edge excitations, the maximum $|\theta|$ and LI increase with amplitude. We also quantify the half width at half maximum (HWHM) of the periodic orbits. As shown in Fig.~\ref{fig:fig3}(b), HWHM slightly increases with amplitude under soft-edge excitation while remains almost constant under rigid-edge excitation. According to both figures, for both soft-edge and rigid-edge excitation, the periodic orbits become more intense as the excitation amplitude increases. However, their spatial localization exhibits a stark, edge-dependent behavior. Soft-edge excitation tends to spatially broaden periodic orbits with increasing amplitude. In sharp contrast, periodic orbits at the rigid edge robustly maintain their high degree of localization, with their spatial width remaining nearly constant across the range of excitation amplitudes. 

Second, the stability of those periodic orbits is theoretically explored. We apply Floquet theory to analyze the evolution of infinitesimal perturbations around a periodic orbit, revealing whether a given periodic motion is robust against small perturbations, or if it is unstable, meaning that small deviations will grow over time. The core of the method lies in calculating the monodromy matrix, which is the linear operator that maps an initial perturbation vector to the perturbation vector after one full period of the orbit. The stability of the periodic orbit is determined by the eigenvalues, $\lambda$, of the monodromy matrix, whose eigenvalues at the time of a full period are known as the Floquet multipliers of the periodic solution. According to Floquet theory, the periodic orbit is linearly unstable if any of its multipliers has a magnitude greater than $1$. If all multipliers have magnitudes less than or equal to $1$, the orbit is linearly (spectrally) stable. The details can be found in Appendix~\ref{appendixC}. By calculating the Floquet multipliers for each periodic orbit found along the continuation path, we find the maximal Floquet multipliers to be practically equal to $1$ regardless of the excitation edge, confirming the spectral stability of the periodic orbits.

After identifying stable periodic orbits through Floquet theory, a natural question arises when our present results~[Fig.~\ref{fig:fig2}(a)] from initial conditions~($\theta=0,\dot{\theta}=0$), lead to drastically different behaviors like delocalization. To understand this phenomenon and investigate the boundary of the basin of bounded motion of stable periodic orbits, we define a straight-line path ($X=\alpha X_{j}, \alpha\in[0,1]$) in the high-dimensional phase space that connects the stable periodic orbits~($\alpha=1$) to the zero solution~($\alpha=0$).

For each point on this straight line, we conduct the simulation from $0$ to $3000T_{b}$ and record the fate of the trajectory. In the soft-edge excitation shown in Fig.~\ref{fig:fig3}(c), for initial conditions close to the periodic orbits which are well within the basin of bounded motion, the simulated trajectory appears to approach the stable periodic orbit. As the initial conditions gradually move to origin~($\alpha=0$), the starting point moves further from the orbit and closer to its basin boundary. At some critical value $\alpha_{c}$, the trajectory will fail to approach the orbit, resulting in a {\it fundamentally} different fate~(delocalization). In comparison, the stability of rigid-edge excitation is displayed in Fig.~\ref{fig:fig3}(d). As the initial conditions move towards the origin, the initial condition is still within the basin of bounded motion. This analysis provides a direct estimate of where the basin boundary lies along this path, explaining why our system starting from the rest state can lead to delocalized dynamics under some excitation amplitudes in the soft-edge excitation. The lying of the initial condition outside of the basin boundary, to the best of our understanding, reconciles the stable nature of the identified  periodic orbits and the apparent emerging dynamical delocalization of the system under the edge excitation.

\section{2D quasiperiodic elliptically geared lattices via synthetic dimension}\label{sec4}
\subsection{Linear spectrum}\label{sec4_1}
\begin{figure}[h!]
    \centering
    \includegraphics[width=0.5\textwidth]{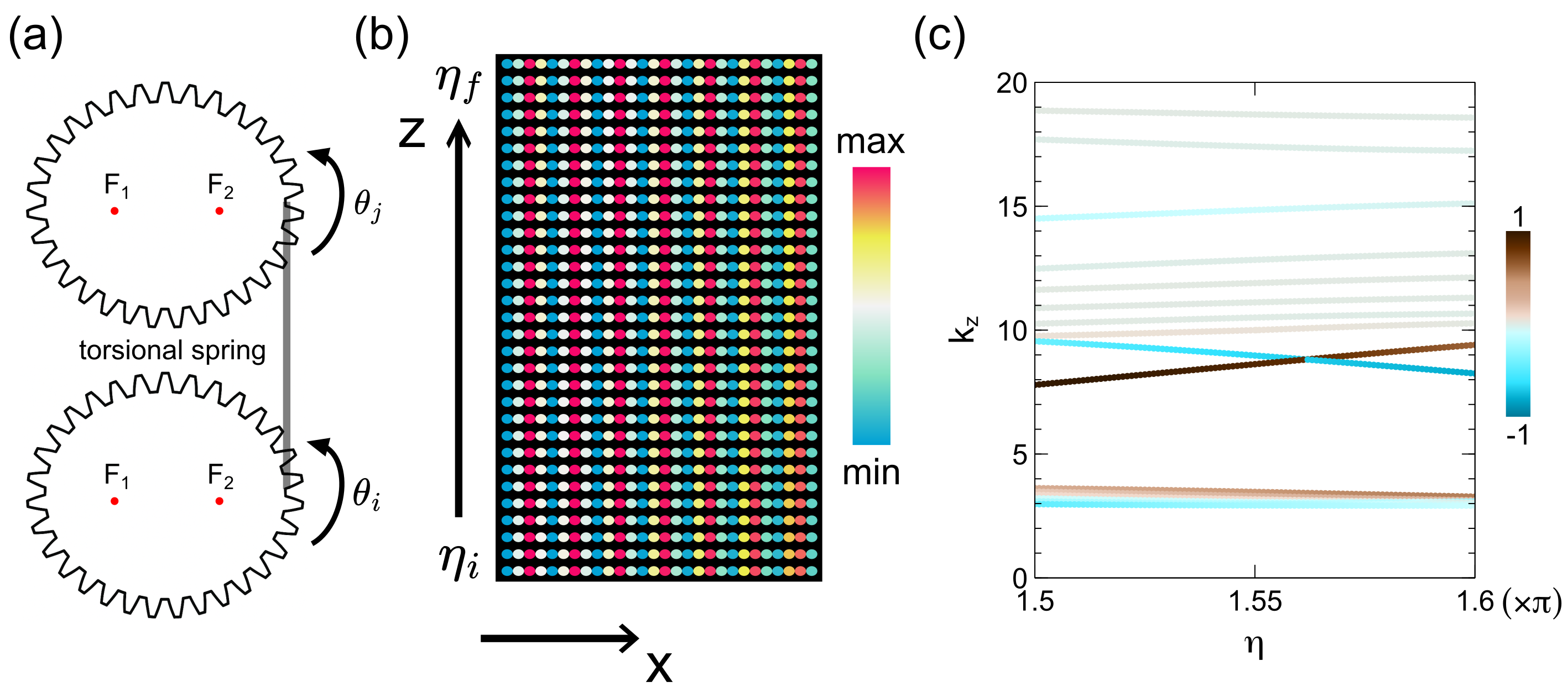}
    \caption{(a) The coupled elliptical gear chains via torsion springs $k_{c}$ along a synthetic dimension $\eta$. $\eta$ varies from $\eta_{i}$ to $\eta_{f}$, i.e., the initial moment of inertial of the first chain is $I_{n}=I_{0}+\mu I_{0}\sin{(2\pi n\xi+\eta_{i})}$ and those of the last chain is $I_{n}=I_{0}+\mu I_{0}\sin{(2\pi n\xi+\eta_{f})}$. The values of the moment of inertia are encoded by the color.
    (b) The wave number $k_{z}$ as a function of $\eta$ for a fixed angular frequency $\omega=2.5$. The color represents the topological states localized at different sides.
    \label{fig:fig4}
    }
\end{figure}

Building on the 1D quasiperiodic chain, we construct a two-dimensional elliptically geared lattice by stacking multiple chains along a second dimension. Each layer corresponds to a given value of the modulation phase $\eta$, effectively sampling the system along the synthetic dimension and producing a corresponding variation in the moment of inertia across the chain. By incrementally varying $\eta$ between layers, we create a synthetic 2D lattice in the $n$ and $\eta$ space. $\eta$ linearly varies from $\eta_{i}=1.53\pi$ to $\eta_{f}=1.59\pi$ across $31$ chains. We couple these layers by introducing additional rotational springs $k_{c}=1$ between corresponding gears in adjacent layers, as illustrated in Fig.~\ref{fig:fig4}(a). These interlayer couplings allow angular motion to propagate along the synthetic dimension, realizing a 2D dynamical system with nearest-neighbor coupling in both real and synthetic directions.

Fig.~\ref{fig:fig4}(b) illustrates a schematic of this 2D synthetic geared lattice. Each colored ellipse represents a gear, with the color denoting the moment of inertia, modulated by the phase $\eta$. To study the dynamical properties of the two-dimensional system, we calculate the spectra. We impose plane-wave harmonic motion along the $z$ direction and the entire spectrum~[Fig.~\ref{fig:fig1}(c)] is shifted by the variation of wave number $k_{z}$, resulting in the spectrum $\omega(\eta,k_{z})$. We then fix the excitation angular frequency $\omega(\eta,k_{z})=2.5$ in the following discussion and generate the relation between $k_{z}$ and $\eta$ shown in Fig.~\ref{fig:fig4}(b). There are two bands featuring topological edge states rendered by color indicating the localization at the soft edge or the rigid edge, which can be represented by the two-level effective Hamiltonian:
\begin{eqnarray}
    \label{equ:equ8}
    \mathcal{H}(\delta\eta)=\begin{pmatrix}
        -\lambda\delta\eta & \Delta k_{z}/2 \\
        \Delta k_{z}/2 & \lambda\delta\eta
    \end{pmatrix},
\end{eqnarray}
where $\lambda$ is a fitting parameter and $\Delta k_{z}$ is the band gap size. Distinct from this common two-band model where a small band gap $\Delta k_{z}$ emerges to enable Landau-Zener transition and topological pumping associated with the condition of adiabaticity~\cite{landau1932theorie,chen2021landau,li2024topological}, our case does not have band gap~($\Delta k_{z}=0$). Therefore, one can expect that even if the number of chains is sufficiently large to enable slow evolution, the final state will be the same as the initial state. While the Landau-Zener transition has been realized in photonics~\cite{xu2023topological}, acoustics~\cite{chen2021landau} and elasticity~\cite{li2024topological}, our finding in preserving topological states along the single edge in the slow evolution remains unexplored before.

\subsection{Nonlinear response}\label{sec4_2}
Similar to behaviors of the 1D counterpart, the 2D system exhibits a significant transition into a nonlinear regime under large excitation amplitudes. Here, geometric nonlinearities disrupt the robust transport of topological states that characterizes the weakly excited system. We systematically investigate how these nonlinear effects in the geared lattice alter its topological states.
\begin{figure}[h!]
    \centering   
    \includegraphics[width=0.48\textwidth]{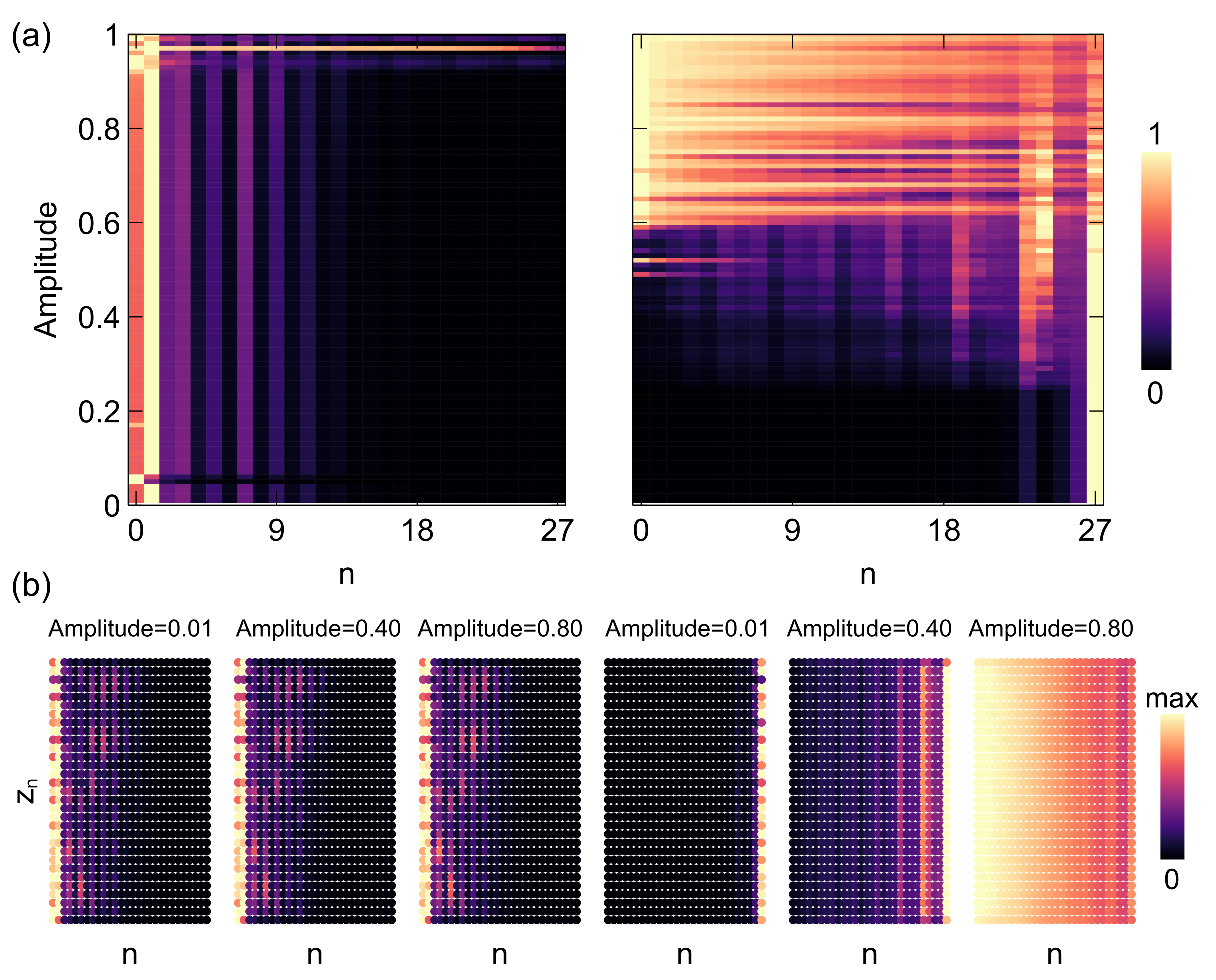}
    \caption{(a) The left panel shows the RMS of $\theta$ when $\eta=1.56\pi$ as a function of excitation amplitude when the excitation is from the soft edge, while the right panel shows that when the excitation is from the rigid edge.
    (b) The first three panels show the RMS of $\theta$ under excitation from the soft edge with amplitudes $0.01$, $0.40$ and $0.80$. The rest panels show the corresponding cases under excitation from the rigid edge.
    \label{fig:fig5}}
\end{figure}

To probe this transition, we apply harmonic excitation to a boundary gear in the first layer. Fig.~\ref{fig:fig5}(a) displays the RMS of rotation angle distribution~($\theta$) for the $16$th gear chain~(middle chain) at $\eta=1.56\pi$ as a function of excitation amplitude. The left and right panels correspond to soft-edge and rigid-edge excitation, respectively. Under soft-edge excitation, topological states propagate effectively at both low and high amplitudes, as evidenced by the edge-localized modes traveling through the physical chain as the synthetic phase $\eta$ evolves~[first three panels of Fig.~\ref{fig:fig5}(b)]. A notable exception occurs near an amplitude of $1.0$, where gears at the soft edge rotate by $\pm\pi$. This behavior contrasts sharply with the 1D system, where topological states are more prone to instability, suggesting that the 2D construction via a synthetic dimension enhances dynamical stability. We visualize the rotation of gears under excitation amplitudes of $0.01$, $0.40$ and $0.80$ in Supplementary Video 3.

In contrast, rigid-edge excitation yields markedly different outcomes. While low-amplitude excitation produces the expected topological mode propagation~[Fig.~\ref{fig:fig5}(b), fourth panel], a significant change occurs as the amplitude increases to $0.40$. The excited waves remain localized at the rigid boundary but also explore the system's configuration space more broadly~[the fifth panel of Fig.~\ref{fig:fig5}(b)], leading to the emergence of highly localized states reminiscent of discrete breathers at specific sites~(e.g., the $16$th, $20$th, and $24$th gears). Upon further increasing the amplitude to $0.80$, the distinct signature of discrete breathers diminishes. Instead, the energy delocalizes, resulting in a more uniform excitation across the bulk of the system~[the sixth panel of Fig.~\ref{fig:fig5}(b)]. The distinction under excitation amplitudes of $0.01$, $0.40$ and $0.80$ is clearly visualized in Supplementary Video 4.

These findings highlight two critical distinctions from the 1D case. First, unlike the 1D system where rigid-edge topological states are robust against amplitude variations, rigid-edge excitation in the 2D system readily triggers a transition to discrete breathers and subsequently to bulk states~[Fig.~\ref{fig:fig5}(a), right panel]. Second, the roles of the excitation boundaries are inverted. In our 2D system, rigid-edge excitation induces delocalization in the nonlinear regime more easily than soft-edge excitation, which is the direct opposite of the behavior observed in the 1D case.
\begin{figure}[h!]
    \centering   
    \includegraphics[width=0.48\textwidth]{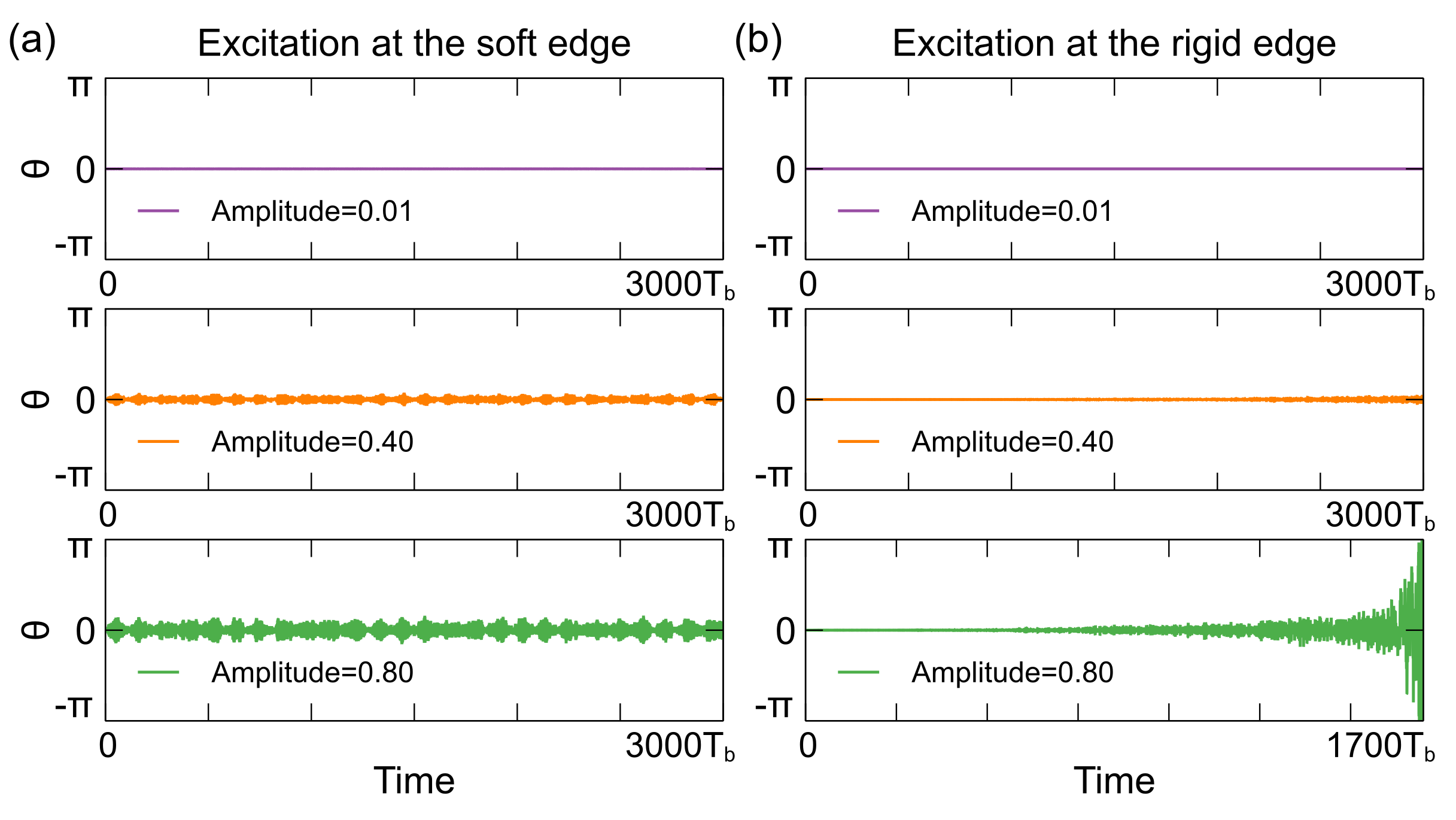}
    \caption{From top to bottom, the evolution of $\theta$ of $\eta=1.56\pi$ and $n=1$ over time for different excitation amplitude $0.01$, $0.40$ and $0.80$ when (a)the excitation is from the soft edge (a) and rigid edge (b).
    \label{fig:fig6}}
\end{figure}

The temporal dynamics shown in Fig.~\ref{fig:fig6} further illustrate the rotation angle of the first gear in the $16$th chain. Under soft-edge excitation, even high amplitudes do not cause a rotational inversion of the gears~[Fig.~\ref{fig:fig6}(a)]. In contrast, as shown in Fig.~\ref{fig:fig6}(b), while low-amplitude ($0.01$ and $0.40$) rigid-edge excitation also avoids inversion, an amplitude of $0.80$ triggers a full rotational inversion, which is induced by a transition wave from the excitation of bulk-like states. This temporal disparity showcases the critical role of boundary-specific nonlinear dynamics in governing topological state stability and transition kinetics.

\section{Conclusion}\label{sec5}
We present a class of mechanical lattices composed of quasiperiodically modulated elliptical gears, where topological mechanics and geometric nonlinearity enable amplitude-controlled topological edge states and nonlinear state transitions~(e.g., from edge to bulk modes). Starting with a 1D chain of elliptical gears, we demonstrate that quasiperiodic modulation of the moment of inertia induces a nontrivial topological phase supporting edge modes, analogous to the AAH model. In the nonlinear regime, this 1D system exhibits amplitude-triggered, edge-dependent transition, where topological edge states delocalize into the bulk under the soft-edge excitation. Extending to a synthetic 2D lattice with modulation phase $\eta$, we achieve edge-state propagation along a single boundary, contrasting traditional adiabatic pumping across chains. Besides, nonlinearity disrupts this 2D topological propagation regardless of excitation edge, inducing complex cross-layer energy redistribution and bulk excitations. Furthermore, to characterize the nonlinear dynamics, we develop a numerical continuation method for tracing periodic orbits across amplitudes and employed Floquet theory to confirm their stability. After we confirm that each periodic orbits solutions are stable, subsequent analysis of basin of bounded motion, by varying initial conditions from stable periodic orbits to the origin, reveals edge-dependent delocalization boundaries.

Our work establishes geared metamaterials as versatile platforms for exploring synthetic dimensions, nonlinear topological phenomena, and amplitude-controlled transport. The observed amplitude-dependent transitions suggest a paradigm for nonlinear topological control, enabling dynamic tuning of mechanical responses~(e.g., energy trapping and signal gating) through excitation amplitude and tailored drives. These findings open avenues for mechanical computation and vibration control. A systematic understanding of the stability of the resulting states as a function of the amplitude of the drive and the corresponding implications for the lattice dynamics pose intriguing questions for future work.

\begin{acknowledgments}
This material is based on the work supported by the National Research Foundation grants funded by the Korea government [Grants No. 2023R1A2C2003705 and No. 2022H1D3A2A03096579 (Brain Pool Plus by the Ministry of Science and ICT)] (J.Y.). This material is also based upon work supported by the U.S. National Science Foundation (Award No. PHY-2110030, PHY-2408988 and DMS-2204702) (P.G.K.). D.Z. thanks the support from National Natural Science Foundation of China (Grant No. 12374157). J.Y. is grateful for the support from the Seoul National University's IAMD and IOER. 
\end{acknowledgments}

\appendix
\section{Equation of motion of the meshed elliptical gears}\label{appendixA}
The Lagrangian of the geared chain and the Euler-Lagrangian equation are written as:
\begin{equation}
    \label{equ:equ9}
    L=T-U=\frac{1}{2}\sum_{n}(I_{n}\dot{\theta}_{n}^2-k_{n}l_{n}^2),
\end{equation}
\begin{equation}
    \label{equ:equ10}
    \frac{d}{dt}\left(\frac{\partial{L}}{\partial{\dot{\theta}_{n}}}\right)-\frac{\partial{L}}{\partial{\theta_{n}}}=0.
\end{equation}
The sliding distance is written as:
\begin{equation}
    \label{equ:equ11}
    \begin{split}
    l_{n}&=s_{e}(\theta_{n})-s_{-e}(\theta_{n+1})\\
    &=a[\theta_{n}-\theta_{n+1}-e_{n}(\sin\theta_{n}+\sin\theta_{n+1})].
    \end{split}
\end{equation}
By substituting Eq.~\eqref{equ:equ11} into $U$, the elastic energy is expressed as:
\begin{equation}
    \label{equ:equ12}
    U=\frac{1}{2}\sum_{n-1}k_{n}a^{2}[\theta_{n}-\theta_{n+1}-e_{n}(\sin\theta_{n}+\sin\theta_{n+1})]^{2}.
\end{equation}
The first term in the Euler-Lagrangian equation:
\begin{equation}
    \label{equ:equ13}
    \frac{d}{dt}\left(\frac{\partial{L}}{\partial{\dot{\theta}_{n}}}\right)=I_{n}\ddot{\theta}_{n}.
\end{equation}
The second term in the Euler-Lagrangian equation:
\begin{equation}
    \label{equ:equ14}
    \begin{split}
        &-\frac{\partial{L}}{\partial{\theta_{n}}}=\frac{\partial{U}}{\partial{\theta_{n}}}\\
        &=k_{n}a^{2}[\theta_{n}-\theta_{n+1}-e_{n}(\sin\theta_{n}+\sin\theta_{n+1})](1-e_{n}\cos\theta_{n})\\
        &-k_{n}a^{2}[\theta_{n-1}-\theta_{n}-e_{n}(\sin\theta_{n-1}+\sin\theta_{n})](1+e_{n}\cos\theta_{n})\\
        &=\tau_{n}-\tau_{n-1},
    \end{split}
\end{equation}
where $\tau_{n}$ is the torque function of gears $n$ and $n+1$ and $\tau_{n-1}$ is the torque function of gears $n-1$ and $n$.

Therefore, the equation of motion of meshed elliptical gear is:
\begin{equation}
    \label{equ:equ15}
    I_{n}\ddot{\theta}_{n}+\tau_{n}-\tau_{n-1}=0.
\end{equation}
We employ the Runge-Kutta method to obtain the time-dependent response of our elliptically geared lattices under the excitation of the harmonic force. The excitation angular frequency is $\omega=2$ for 1D case and $\omega=2.5$ for 2D case. The excitation amplitudes are listed in the main text.

For the spectra calculation, we linearize the equation of motion by considering the small rotation angle~(i.e., $\theta\rightarrow0$, thus, $\sin\theta\rightarrow\theta$ and $\cos\theta\rightarrow1$). The linearized equation of motion is expressed as:
\begin{equation}
    \label{equ:equ16}
    \begin{split}
        I_{n}\ddot{\theta}_{n}&+k_{n}a^{2}(\theta_{n}-\theta_{n+1}-e_{n}\theta_{n}-e_{n}\theta_{n+1})(1-e_{n})\\
        &-k_{n}a^{2}(\theta_{n-1}-\theta_{n}-e_{n}\theta_{n-1}-e_{n}\theta_{n})(1+e_{n})=0.
    \end{split}
\end{equation}
Eq.~\eqref{equ:equ16} is then organized to show the coefficients in front of the variables:
\begin{equation}
    \label{equ:equ17}
    \begin{split}
        I_{n}\ddot{\theta}_{n}&+k_{n}a^{2}[(1-e_{n})^{2}+(1+e_{n})^{2}]\theta_{n}\\
        &-k_{n}a^{2}(1+e_{n})(1-e_{n})\theta_{n-1}\\
        &-k_{n}a^{2}(1+e_{n})(1-e_{n})\theta_{n+1}=0.
    \end{split}
\end{equation}
Therefore, the coefficients in Eq.~\eqref{equ:equ4} correspond to $\alpha=k_{n}a^{2}[(1-e_{n})^{2}+(1+e_{n})^{2}]$, $\beta=-k_{n}a^{2}(1+e_{n})(1-e_{n})$ and $\gamma=-k_{n}a^{2}(1+e_{n})(1-e_{n})$. The external driving torque $\tau_{ext}$ is in the form of $A\cos{\omega t}$, where $A$ is the driving amplitude and $\omega$ is the driving angular frequency.

Based on this linearized equation of motion, the spectrum of our gear system $\omega$ can be obtained by solving the eigenvalue equation:
\begin{equation}
    \label{equ:18}
    (-\omega^{2}\bm{I}+\bm{D})\bm{\Theta}=0.
\end{equation}
Here, $\omega$ denotes the angular frequency. $\bm{I}$ denotes the lumped matrix of moment of inertia containing the moment of inertia $I_{n}$ of each gear. $\bm{D}$ is the dynamical matrix that can be constructed by $\alpha$, $\beta$ and $\gamma$. $\bm{\Theta}$ is the corresponding eigenmode $[\theta_{1},\theta_{2},...,\theta_{n}]^{T}$. $\omega$ is then solved under different $\xi$, $\eta$ and $k_{z}$ depending on different scenarios.

\section{Continuation method to find periodic orbits}\label{appendixB}
The primary method we have detailed for finding periodic orbits in our driven system is based on finding a fixed point of the Poincar\'e map in phase space. The goal is to find the specific initial state vector $X_{0} = [\theta,\dot{\theta}]^{T}$, which, under the system's dynamics, repeats itself after one period of the external drive. For a driven system with period $T_{b}$, the Poincar\'e map, $\mathcal{P}$, evolves the state vector over one period: $X(T_{b})=\mathcal{P}(X_{0},A)$, where $A$ is the excitation amplitude. A periodic orbit must satisfy the fixed point condition: $X_{0}=\mathcal{P}(X_{0},A)$. This is formulated as a root-finding problem for the function $G$: $G(X_{0},A)=X_{0}-\mathcal{P}(X_{0},A)=0$. We adopt the Newton iterative method to solve the equation, where the general form of the update rule from a current guess $X_{k}$ to the next guess $X_{k+1}$ is:
\begin{equation}
    \label{equ:19}
    X_{k+1}=X_{k}-[J_{G}(X_{k})]^{-1}G(X_{k}),
\end{equation}
where $J_{G}=\frac{\partial G}{\partial X_{0}}=I-\frac{\partial\mathcal{P}}{\partial X_{0}}=I-M$, with $M$ being the monodromy matrix.

We start from the topological state in the neighborhood of the linear regime~(specifically, for $A=0.001$) and use one of the time-dependent solution after the transient stage as the initial guess $X_{0}$. Then, the numerical continuation process begins by finding the solution at a slightly larger excitation amplitude. This step-by-step process allows one to follow a branch of periodic solutions $X_{j}$ as the amplitude increases.

\section{Stability analysis of periodic orbits}\label{appendixC}
To determine if an identified periodic orbit is physically stable, its linear stability is analyzed using Floquet theory. This involves linearizing the equations of motion around the periodic orbit to study the evolution of infinitesimal perturbations. The process begins by considering a perturbed solution $\psi_{j}(t)=X_{j}(t)+\varphi_{j}(t)$ around $X_{j}$, which leads to a linear differential equation for the perturbation $\varphi_{j}(t)$. This equation is then expressed in a first-order matrix form:
\begin{equation}
    \label{equ:20}
    \dot{\Omega}(t)=J_{\Omega}(t)\Omega,
\end{equation}
where $\Omega(t)$ is the state vector of the perturbations and $J_{\Omega}(t)$ is the system's Jacobian matrix evaluated along the periodic orbit.

The stability is determined by the monodromy matrix, $M$, which maps an initial perturbation $\Omega(0)$ to the perturbation after one full period: $\Omega(T_{b})=M\Omega(0)$. The monodromy matrix is calculated by finding the $n \times n$-dimensional state transition matrix $\Phi(t)$ which is the solution to the matrix differential equation:
\begin{equation}
    \label{equ:21}
    \dot{\Phi}(t)=J_{\Omega}(t)\Phi(t),
\end{equation}
with $\Phi(0)=I$. The monodromy matrix is given by $M=\Phi(T_{b})$. The eigenvalues $\lambda$ of $M$ are the Floquet multipliers. The periodic orbit is linearly unstable if any Floquet multiplier has a magnitude greater than $1$. Otherwise, it is linearly (spectrally) stable.

\bibliography{references}

\end{document}